# Detecting depression in dyadic conversations with multimodal narratives and visualizations

Joshua Y. Kim[1], Greyson Y. Kim[2], and Kalina Yacef[1]

[1] University of Sydney, Darlington NSW 2006, Australia
[2] Success Beyond Pain. Success WA 6164, Australia
{josh.kim,kalina.yacef}@sydney.edu.au
greyson.kim@gmail.com

**Abstract.** Conversations contain a wide spectrum of multimodal information that gives us hints about the emotions and moods of the speaker. In this paper, we developed a system that supports humans to analyze conversations. Our main contribution is the identification of appropriate multimodal features and the integration of such features into verbatim conversation transcripts. We demonstrate the ability of our system to take in a wide range of multimodal information and automatically generated a prediction score for the depression state of the individual. Our experiments showed that this approach yielded better performance than the baseline model. Furthermore, the multimodal narrative approach makes it easy to integrate learnings from other disciplines, such as conversational analysis and psychology. Lastly, this interdisciplinary and automated approach is a step towards emulating how practitioners record the course of treatment as well as emulating how conversational analysts have been analyzing conversations by hand.

## 1    Introduction

When people speak, a lot of information is communicated at several levels. The content, as well as the way the speech is delivered, gives us hints about the emotions and moods of the speaker. To be a good listener in the conversation or a third-party conversation analyst, one must consider a wide range of information, paying attention to the choice of words, attitudes, and emotions, for example.

Analyzing conversations is complex and time-consuming. In specialized settings, such as a clinical psychology setting, expert insights are required to get a sense of what to look out for – both in terms of what was said and how it was said. For example, people who are clinically depressed tend to lose interest in things they were previously interested in [1] (which relates to what was said). Also, while responding, the clients tend to have a longer response time and/or have reduced affect display in facial expression [2, 3] (which relates to how it was being said). As an anecdotal example from one of our authors' experience as a practicing clinical psychologist, his experience is aligned to the aforementioned literature and he also found that during the initial treatment, depressed clients tend to employ avoidance coping, such as responding with ambivalence (e.g., I don't know') when they are asked to encounter their inner experience.

In another related discipline, conversational analysis, multimodal information from dialogs have been transcribed using a technical system developed by Jefferson [4]. This system encodes information about how it was being said in addition to what was being said. The challenge is that the manual encoding of such information is very time-consuming and susceptible to human error.

In this paper, we alleviate the manual-intensive problem of representing and visualizing multimodal information. We customize our conversational analysis system so that it takes in temporal multimodal information, weaves them into the verbatim transcript and automatically generates a prediction score for the depression state of the individual. We discuss insights from current practices in the clinical psychology industry to help inform our construction of the multimodal narrative. With the help of current artificial intelligence (A.I.) algorithms, we extract features to accomplish a task that demands both attention and domain-specific expertise. The experiments show that our proposed method performs better than the baseline model. Lastly, we present an example of how the content of a conversation can be visualized and analyzed intuitively by humans.

## 2    Related Works

### 2.1    Textual multimodal representation

Representations of multimodal information are typically vector-based, i.e., numerical. However, we are proposing a textual form of multimodal representation. There are two closely related tasks in the domain of machine learning – visual question-answering (VQA) [5] and dense video captioning [6] tasks. In these tasks, the model is trained to take in the video input and output a sequence of text that describes the video or answer questions. Our objective differs from these tasks, because our generated text is both used as an intermediary step for downstream models as well as a final product to be used in visualizations. Therefore, our objectives surrounding the generated text are (1) interpretability of the downstream model, and (2) whether the downstream model would be performant. Consequently, we are not concerned with the measures typically used in VQA and dense video captioning tasks to compare against the ground truth. Instead, our objective measure would be on the downstream model performance.

### 2.2    Detecting depression with automated conversational analysis

Recently, there has been a growing interest in interdisciplinary research that assesses human conversations automatically. We focus on applications that detect depression and discuss past attempts at detecting depression in conversations. The Audio/Visual Emotion Challenge (AVEC), in the year of 2016, 2017 and 2019, has invited researchers to predict the level of depression severity (measured by the PHQ-8 questionnaire) from audio-visual recordings of a clinical interview. Since our dataset was also used in the AVECs, past attempts in the AVECs are highly relevant.

Although the challenges were about analyzing conversations, we did not find any past accepted papers that were concerned with the textual presentation of the analysis. Many accepted papers did, however, use text as an input. Some used the verbatim transcript in the text format as input into the model. Others extracted numeric features from the transcripts in three main ways – timings, keyword searches, and emotional indices.

## 3  Data

### 3.1  Data collection

We used a publicly available multimodal dataset, Distress Analysis Interview Corpus [7]. It contains interviews of individuals conducted by a virtual human designed to help diagnosis of psychological distress conditions. Self-reported PHQ-8 scores are provided as the dependent variable, with an interval scale from 0 to 24, and larger scores indicate greater severity. In total, there are 219 participants. This dataset was also used in the AVEC 2019 challenge. The organizers have performed the digitization step to extract baseline features [8]. The scripts that replicate the baseline features extraction is publicly available[1]. Of the baseline features, we used the Geneva Minimalistic Acoustic Parameter Set (egemaps) and Facial Action Units (AU).

### 3.2  Feature extraction

Using the provided data and features, we extracted more features and classified them into three different levels of inputs. Fig. 1 provides an overview of the prediction process and the distribution of the PHQ score. The baseline level of input is the session-level numeric features, followed by a session-level coarse summary. Lastly, the multimodal narrative comprised of both the verbatim transcript and multimodal information at the talk-turn level. The motivation is to produce an interpretable set of features that considers the temporal nature of conversational analyses. In total, there are nine features extracted at the session-level and four features extracted at the talk-turn level.

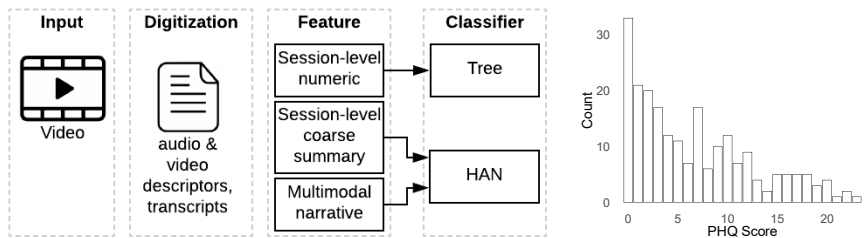

**Fig. 1.** The high-level architecture of the pipeline (left) and distribution of PHQ score (right).

### 3.2.1 Session-level numeric features

We created three families of session-level inputs – *prosody*, *actions*, and *demographics*. Table 1 gives an overview of all features. Summary statistics refer to the maximum, minimum, average, and standard deviation of the feature for the entire session.

---

[1] https://github.com/AudioVisualEmotionChallenge/AVEC2019

**Table 1.** Session-level numeric features.

| Family | Child | Feature |
|---|---|---|
| Demo-graphics | Talkativeness | Total word count & Total distinct word count |
| | Big 5 Personality | Percentile scores for each of the big 5 personality |
| | Gender | Predicted Male/Female |
| Actions | Laughter | Total laughter count |
| | Facial expression | Summary statistics of the given AU intensity and count values |
| Prosody | Delay | Summary statistics of time gaps between talk-turns |
| | Speech rate | Average speech rate |

Firstly, for the family of *demographics*, we investigated the effects of *talkativeness*, *Big 5 Personality*, *Gender*. *Talkativeness* is selected because depressed individuals have been found to show reduced response length and poor articulation of distinct words [9].

Of the *Big 5 Personality*, higher neuroticism and lower conscientiousness have been found to correlate with higher depression. In a meta-analysis that reviewed 851 effect sizes based on 175 articles, strong effect sizes have been observed in the correlations between depression and neuroticism ($d = 1.33$), and conscientiousness ($d = -.90$) [10]. Interestingly, it has been found that the *Big 5 Personality* did not correlate with psychomotor retardation, while the *Big 5 Personality* correlates with the negative affect component in depression [11]. Hence, the present study included proxies of depression from both affective (*Big 5 Personality*) and physiological dimensions (*Speech Delay*, *Speech Rate*, *Talkativeness*).

Gender differences exist in the manifestation of depression. Firstly, in a large study using about 80,000 individuals, women tend to report more depressive symptoms than men across all age groups [12]. Secondly, in perpetuating depression, chronic strain, low mastery, and rumination were reported to be more common in women than in men [13]. Therefore, the factor of gender is included to capture these possible systematic differences in depression manifestation.

The method of extracting features for the family of *demographics* is as follows. *Talkativeness* is measured by the total word count and total distinct word count from the supplied transcripts. *Big 5 Personality* is computed using the IBM Watson Personality Insights API, which returns the five percentile scores – one for each *Big 5 personality* – which are used as inputs. *Gender* is predicted by a regularized logistic regression model using the egemaps features supplied. Using all 16 egemaps variables, we obtained an accuracy of 99.1% on the pooled predictions from the cross-validation. We used one-hot encoding to represent gender as a numeric feature.

Secondly, for the family of *actions*, we investigated the effects of *laughter* and *facial expressions*. The reduction in *laughter* frequency has been found to be a symptom of depression, which may be linked to humor deficit and increased feelings of anhedonia observed in depressed patients [14]. *Facial expressions* of emotions are also expected to be reduced with depression [15].

The method of extracting features for the family of actions is as follows. *Laughter* events were detected using the open-source algorithm [16]. We summed up the total number of laughter events per session. As for *facial expressions*, we computed summary statistics for the intensity and count values of each AU. To limit the number of input features, we only computed summary statistics for four AU (AU5, 17, 20, 25) because these have been found to be effective in Yang et al. [17]. The description of each of the four AUs are – AU5: Upper lid Raiser; AU17: Chin raiser; AU20: Lip stretcher; AU25: Jaw drop [18].

Lastly, for the family of *prosody*, we investigated the effects of delay and speech rate. These two features were selected because of the existing literature that suggests the following. Firstly, higher delays in responses (also known as speech pause time) have been found to positively correlate with higher depression scores [2, 3]. Secondly, a lower speech rate has also been found to also correlate with higher depression scores [19].

The method of extracting features for the family of *prosody* is as follows. We computed the summary statistics of time gaps between talk-turns. As for speech rate, the overall average speech rate is computed by dividing the total number of words spoken by the total number minutes of talk-turn duration.

### 3.2.2 Session-level coarse summary

In addition to numerical inputs, we investigated whether it is possible to represent the inputs in the form of text. The motivation is that in clinical psychology consultations, practitioners are ethically obligated to summarize session-level progress details, to inform subsequent assessments and interventions in the course of treatment (see https://www.psychology.org.au for example). This investigation is a step towards emulating how practitioners record the course of treatment and provides insights on analyzing the session-level coarse summaries. In this section, we first explain the automatic generation of a text representation, then we motivate the extraction of a new family of input features – *comprehension*. The overview of templates for the session-level coarse summary is provided in Table 2.

**Table 2.** Templates for the session-level coarse summary.

| Family | Child | ID | Template |
|---|---|---|---|
| Demo-graphics | Talkativeness | 1 | 'number of words **high** number of distinct words **high**' |
| | Big 5 Personality | 2 | 'openness **very high**' |
| | Gender | 3 | The participant is **female**. |
| Actions | Laughter | 4 | 'laughter counts **high**' |
| | AU | 5 | 'minimum lip depressor **very low** maximum lip depressor **low** average lip depressor **low** variance lip depressor **low**' |
| Prosody | Delay | 6 | 'minimum delay **very low** maximum delay **low** average delay **low** variance delay **low**' |
| | Speech rate | 7 | 'speech rate **high**' |
| Compre-hension | Yang et al. [17] | 8 | See Table 3 |
| | DSM-5 | 9 | See Table 3 |

The method of converting session-level features into text representation involves standardization of the numerical inputs across the training dataset. For each of the numerical inputs, we computed the standardized score (z-score) using the mean and standard deviation obtained from the training fold. Depending on the standardized score, we inserted templates 1 to 7 (except 3), where possible values in the bold-face were "very low" (z-score < -2), "low" (z-score < -1), "high" (z-score > 1) and "very high" (z-score > 2). We did not insert any templates when the z-score was within the normal range of -1 to 1 to keep the narrative succinct. As for template 3, we used the predicted gender from the logistic regression model.

Since we are changing the representation of the session from numerical features to text features, we also investigated the addition of a new family of input features – *comprehension*. The recent progress of machine comprehension has prompted the research question of the benefits of using trained machine comprehension models to extract text features for downstream supervised learning. Machine comprehension models take in two inputs – the question and the passage – and return an extracted phrase from the passage that is most related to the question. In our application, we used the model as a targeted summarization tool. The *comprehension* input family consisted of two sets of questions which we discuss in turn. To extract the answers, we used the pre-trained Bidirectional Attention Flow model from AllenNLP [20]. We also set the minimum probability threshold to be 0.1, so that answers that are deemed to be low-confidence by the model are replaced by "not applicable".

The first set of *comprehension* questions were derived from Yang et al. [17], where the authors conducted a content analysis of the transcripts using keywords-matching to identify whether the participant is (a) previously diagnosed, (b) sleeping well, (c) shy or outgoing, (d) feeling bad or good. The second set of *comprehension* questions were derived from the DSM-5 [1]. The DSM-5 is widely used in Australia by psychologists and psychiatrists to make a diagnosis of depression. We present the set of derived questions along with a sample answer extracted from the transcripts in Table 3.

**Table 3.** Sample *comprehension* features derived.

| Yang et al [17] | | DSM-5 [1] | |
|---|---|---|---|
| **Question** | **Sample answer** | **Question** | **Sample answer** |
| Am I diagnosed? | no I never been formally diagnosed | Do I feel depressed most of the day? | no lately I've still been pretty depressed |
| Am I sleeping well? | I have not been sleeping | Do I lose interest? | I was not interested in things that are that I would normally interested in and you know I was sort of withdrawn |
| Am I shy? | I'm not I'm not extremely shy | Do I feel tired? | fatigued I'm very tired |
| How am I feeling lately? | irritated tired lazy | Do I feel worthless? | I feel like invisible |
| | | Do I feel like dying? | I felt like I couldn't cope |

### 3.2.3 Multimodal narrative

In this section, we also investigated the effects of adding the entire transcripts, annotated with multimodal information. In effect, we added information on *what* was being said as well as *how* it was being said at a talk-turn level.

At the talk-turn level, we introduced two families of information – *prosody*, and *actions*. We weaved this information into the transcript to create the multimodal narrative. Table 4 gives an overview of the templates, the bold-face indicates a variable.

**Table 4.** Templates for the multimodal annotations

| Family | Child | ID | Template |
|--------|-------|-----|----------|
| Actions | Laughter | 10 | 'the participant **laughed and** said' |
| Prosody | Delay | 11 | 'after **two** hundred milliseconds' |
| | | 12 | 'a **long** delay' |
| | Speech rate | 13 | '**quickly** said' |

The laughter detection algorithm returns the time window of the detected laughter. Using this time window, we inserted template 10 if the window of the laughter is completely contained within the time window of the talk-turn.

The method of computing delay is through the provided time window for each talk-turn in the transcripts. The nominal values (template 11) of delay could be used by the model to compare the delay lengths across sessions; however the drawback of is that it does not help the model consider the within-session variation in delay. Therefore, to tackle this problem, we added template 12 where the template is dependent on the standardized duration of the delay "long" ($1 \leq$ z-score $< 2$) and "significantly long" ($2 \leq$ z-score). The mean and standard deviation of the delay are calculated using all talk-turn delays from the same session. Lastly, we appended template 13 to annotate speech rate variation within the session. Similar to template 12, the possible values of template 13 are dependent on the standardized words per minute for each talk-turn – "very slowly" (z-score $< -2$), "slowly" (z-score $< -1$), "quickly" (z-score $> 1$), and "very quickly" (z-score $> 2$).

## 4 Analysis

### 4.1 Evaluation Metric

To evaluate the performance of the models, the Concordance Correlation Coefficient (CCC) was used, this was also the metric used in the official AVEC 2019 challenges [8]. The CCC is defined by:

$$r_c = \frac{2r\sigma_x\sigma_y}{\sigma_x^2 + \sigma_y^2 + (\mu_x - \mu_y)^2} \qquad (1)$$

where *r* represents the *Pearson's correlation coefficient (PCC)* between two vectors (ground-truth and prediction), μ represents the mean of each vector, and σ represents the standard deviation of each vector.

## 4.2 Ablation test design

To investigate the benefits of including different families of input features, and different learning algorithms, we designed a series of ablation tests. We discuss the motivations for our ablation test design, as illustrated in Fig. 2. The dotted boxes correspond to the granularity of feature input, as described in Fig. 1. The grey boxes denote the learning algorithm. The white boxes denote the families of inputs that were common to our existing system; whilst the blue boxes denote the families of inputs that are newly added features in this paper. For each model, we start from the left, which uses only *demographic* features. The ablation test setup is such that we add one input family at a time, starting with A, then P and others. We discuss the four motivations for this design.

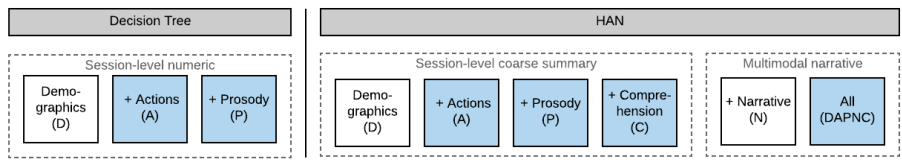

**Fig. 2.** The high-level architecture of the prediction pipeline.

The first motivation is the ability to compare the differences in performance between the two algorithms, keeping the input configurations similar. Therefore, in Fig. 2, we observe that it is possible to compare the performance of the decision tree versus the HAN for the first three configurations (D, DA, DAP).

The second motivation is the ability to investigate the additional benefits of progressively adding families of input into the configuration. Therefore, we have elected to add one family at a time. We have chosen demographics to be the starting configuration because this family of inputs is always populated for text inputs. In other words, because the other two families (*actions* and *prosody*) are only populated when they contain high standardized values (see section 3.2.2), they are not suitable as the starting configuration.

The third motivation is the ability to investigate the benefits of employing the pretrained machine comprehension model. Therefore, we have introduced a split at the end of the DAP configuration. On one path, we used the machine comprehension model to perform targeted summarization. On the other path, we used our automatically generated multimodal narrative, which includes both the multimodal information and verbatim transcript.

Lastly, the fourth motivation is the ability to investigate the effects of using all extracted information in the form of text in the DAPNC configuration.

## 4.3 Regression tree

The regression tree used is from the rpart package within R. We used the "anova" splitting method to train the regression tree. We tuned the hyperparameters minimum split, maximum depth, and cp through a grid search and the cross-validation. We report in Table 5 the performance metrics using different inputs. The cross-validation is performed on the entire dataset.

### 4.4 HAN

With the text features, each word is represented by the set of Glove word embeddings (300-dimensions). We tuned the learning rate, number of GRU units, recurrent dropout, GRU dropout, and L2 regularization via random-search. We used the Stochastic Gradient Descent optimizer. Batch size is set to 8. The training process consists of 350 epochs, with no early stopping. We have also clipped the range of the predicted values to be between 0 and 24. The performances of HAN are also presented in Table 5. To facilitate all pairwise comparisons, we also computed whether the performance differences are statistically significant through bootstrapping the CCC differences (N=1000). At the 0.95 significance level, we found significant differences in two pairs - Tree-D vs. HAN-DAPN, and Tree-D vs. HAN-DAPCN. At the 0.90 significance level, we found a significant difference in one additional pair – Tree-D vs. HAN-DAPC.

**Table 5.** Cross-validation results for regression tree and HAN. The standard deviation of performance is presented in brackets.

|  | Regression Tree | HAN |
|---|---|---|
| Family | CCC | CCC |
| D | 0.144 (0.065) | 0.171 (0.027) |
| DA | 0.226 (0.125) | 0.200 (0.035) |
| DAP | 0.234 (0.052) | 0.239 (0.056) |
| DAPC |  | 0.291 (0.118) |
| DAPN | N/A | 0.297 (0.132) |
| DAPNC |  | 0.302 (0.092) |

## 5 Discussion

For regression tree, we observe that the CCC cross-validation performance increases significantly from the D configuration (CCC=0.144) to the DA configuration (CCC=0.226) and then followed by a small increase in the DAP configuration (CCC=0.234).

As for the HAN, the increase of CCC is almost linear as we progressively add the session-level inputs – from the D (CCC=0.171), to DA (CCC=0.200), and then the DAP configuration (CCC=0.239). When we added the comprehension features, the performance increases markedly (CCC=0.291). After that, the performance of the DAPC configuration is similar to DAPN (CCC=0.297) and the configuration of DAPNC, which contains all text features, is the best (CCC=0.302).

The DAPC configuration is an efficient way to extract text features for this supervised learning task. This is evident from two observations. Firstly, the performance increase from the DAP to DAPC configuration is high. Secondly, the difference between the DAPC and DAPN configuration is small. However, we posit that the challenge here is asking the machine comprehension model the right questions, which we attempted to overcome by using past research, including the DSM-5.

When we hold the input configuration constant, we found that text representation had the same level of performance as the regression tree when representing session-level inputs (DAP configuration). The CCC performances of other participants of the AVEC 2019 challenge – using the same number of observations and performance metric (CCC) – are not published at the time of writing this paper.

## 6    HAN Visualization

In the closing section, we demonstrate how the HAN model could be used to support humans analyze conversations, through using the attention weights. We used both the talk-turn as well as word-level attention weights to construct the visualization manually. Here, a talk-turn represents a question (answer) to (from) the machine comprehension model. We extracted and standardized attention weights from all talk-turns within the same session. Having computed the standardized talk-turn-level and word-level weights, we use the following style schemes to construct the visualization.

| z-score | Importance | Word Color | Size | Talk-turn Color | Label |
|---------|-----------|------------|------|-----------------|-------|
| < 0 | Nil | | +0 | | N |
| < 1 | Low | | +1 | | L |
| < 2 | Medium | | +2 | | M |
| ≥ 2 | High | | +3 | | H |

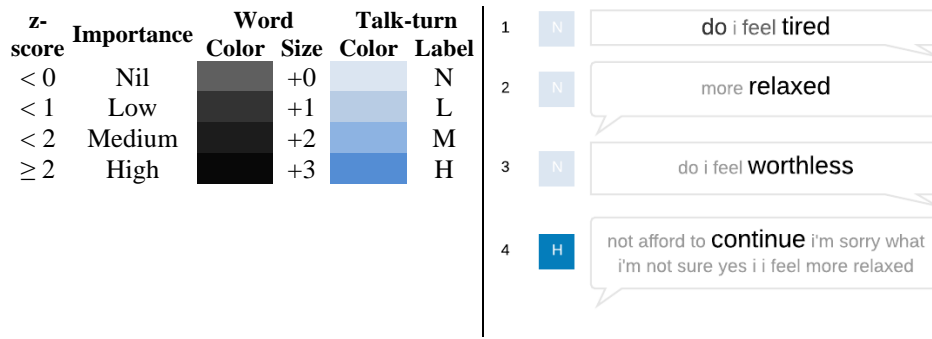

**Fig. 3.** Conversation analysis example.

In the visualization, the numbers on the left indicate the talk-turn number. The analyst could first get a sense of the relative importance of all talk-turns by looking at the talk-turn-level attention weights. Then, the analyst could dive deeper and analyze the relative importance of each word.

We picked an example that has the actual PHQ score of 20, and the model predicted a score of 15. In Fig. 3, we illustrate a short extract from the DAPC configuration that has talk-turns with high attention weights. In the extract, we observed that only talk-turn 4 are having high importance in making the prediction score of 15. More interestingly, we noted that the words marked as high importance are concepts semantically related to depression.

## 7    Conclusion

In this paper, experiments on the DAIC-WOZ dataset [7] were presented, with a focus on the multimodal text representation. Our proposed approach constructs a text narrative and then uses it as input to a model. Finally, both the model and the narrative are used to construct a visualization to analyze the dialog.

We have demonstrated that this approach could accommodate customizations, allowing the researcher to inject research findings from cross-disciplinary literature. In our application, we injected *talkativeness*, *big 5 personality*, *laughter*, *delay*, *speech rate*, *comprehension* information into our multimodal narrative. These selected features are guided by the psychology as well as the conversational analysis literature.

The ablation tests showed that text representations perform just as well as the regression trees. However, the flexibility of text representation makes it possible to extract more information through the pre-trained machine comprehension model and achieve a cross-validation CCC of 0.291. Using all the information, our system achieves a cross-validation CCC of 0.302. For future research directions, we would like to better understand the effectiveness of the produced visualizations and text summaries as well as the time savings from this automated approach through user-studies. There is also future work to improve or widen the multimodal feature extraction to further improve the performance of HAN.

Lastly, the visualization is a step towards emulating the way human conversation analysts analyze conversations, and a step towards emulating how practitioners keep a text record of the course of treatment.